\documentclass[runningheads]{llncs}
\usepackage[T1]{fontenc}
\usepackage{graphicx}
\usepackage{booktabs}
\begin{document}
\title{Love in Action: Gamifying Public Video Cameras for Fostering Social Relationships in Real World}
\titlerunning{Love in Action: Gamifying Public Video Cameras}
%
\author{Zhang Zhang\inst{1,2}\orcidID{0000-0001-9425-3065} \and
Da Li\inst{1} \and
Geng Wu\inst{1} \and
Yaoning Li\inst{1} \and
Xiaobing Sun\inst{1} \and
Liang Wang\inst{1,2}}
%
\institute{NLPR, MAIS, Institute of Automation, Chinese Academy of Sciences (CASIA) \and
School of Artificial Intelligence, University of Chinese Academy of Sciences (UCAS) \\
\email{zzhang@nlpr.ia.ac.cn}, \email{\{da.li, geng.wu, yaoning.li xiaobing.sun\}@cripac.ia.ac.cn}, \email{wangliang@nlpr.ia.ac.cn}}
%
\maketitle              
\begin{abstract}
In this paper, we create "Love in Action" (LIA), a body language-based social game utilizing video cameras installed in public spaces to enhance social relationships in real-world. In the game, participants assume dual roles, i.e., requesters, who issue social requests, and performers, who respond social requests through performing specified body languages. To mediate the communication between participants, we build an AI-enhanced video analysis system incorporating multiple visual analysis modules like person detection, attribute recognition, and action recognition, to assess the performer's body language quality. A two-week field study involving 27 participants shows significant improvements in their social friendships, as indicated by Self-reported questionnaires. Moreover, user experiences are investigated to highlight the potential of public video cameras as a novel communication medium for socializing in public spaces.

\end{abstract}
\section{Introduction}
Human beings are social creatures by nature. Meaningful social relationships are crucial for individual health and well-being, as well as for social cohesion in communities and societies \cite{Xavier_PHD}. However, in today's fast-paced world, individuals often experience persistent stress or anxiety due to various pressures. In the workplace, some people primarily focus on their own Key Performance Indicators (KPIs) and interacts with others in a utilitarian manner, lacking genuine empathy for others. After work, many people opt to retreat to the comfort of their homes, spending time with televisions, computers and smartphones, immersing themselves in video games or watching content on social platforms like Youtube and Tiktok. Despite social medias facilitate users to be connected with the same interests, a study \cite{Melissa18} has found that high usage of social media can actually increase feelings of loneliness rather than alleviate them. Other research \cite{Primack17} also suggests that online interactions cannot be substituted for real social relationships, and socially isolated individuals often tend to spend more time on social medias. Consequently, the development of in-person relationships in the real world cannot be completely replaced by online relationships.

To facilitate in-person relationships in the real world, researchers have investigated various Human-Computer Interaction (HCI) technologies and gamification designs, such as live-streaming services \cite{Sheng20}, eSports games \cite{guo17}, and Location-Based Games (LBGs) \cite{Xavier21}. Notably, LBGs like Pokémon GO \cite{pokemon16} have proved effective to develop in-person relationships \cite{bhattacharya19} \cite{pokemon17} and bring positive experiences of sociability \cite{Xavier21}. Typically, LBGs use Global Positioning System (GPS) technology to perceive and connect individuals. Different from previous LBG studies, this work leverages an advanced AI technique, Intelligent Video Surveillance (IVS), to create a LBG with richer visual perceptions of players, and evaluates its influence on fostering in-person relationships.

In addition to incorporating a new positioning technique in LBGs, we are also motivated by three factors to investigate the viability of video surveillance systems as an innovative social platform.
\begin{figure}[tb]
  \centering
  \includegraphics[width=\linewidth]{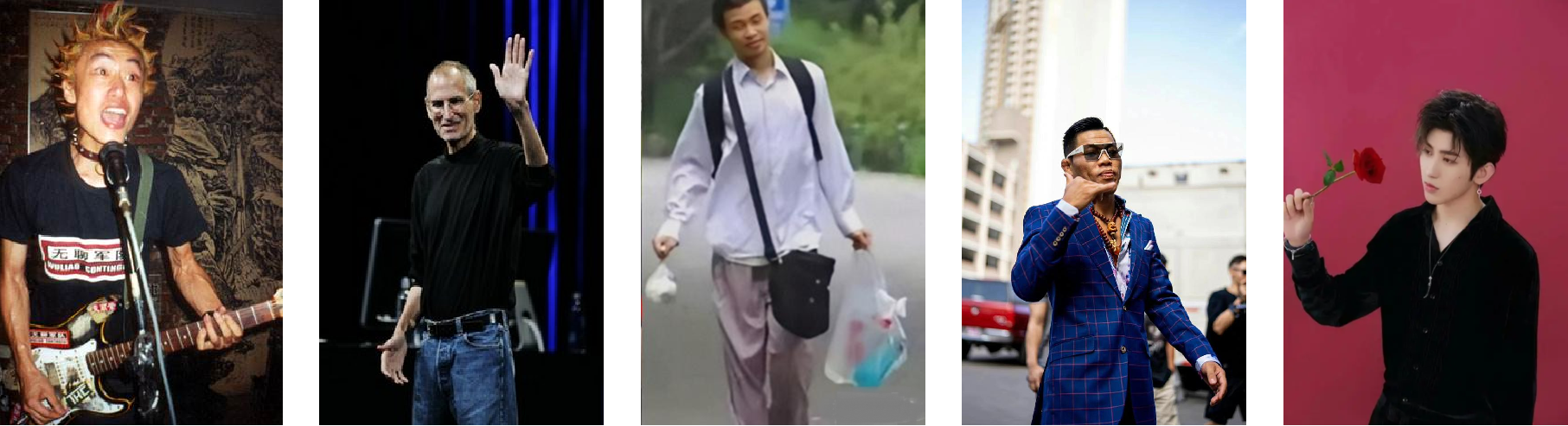}
  \caption{Some representative self-expressions using different clothing fashions, accessories and postures, which are termed \emph{body languages} in this paper. The images are sourced from online news.}
  \label{fig:style}
\end{figure}
\begin{itemize}
    \item Video surveillance originally derives from the concept of panopticon proposed by Jeremy Bentham (1748-1832), who designed a panoptic society transparent to itself where people would finally deter delinquency and keep morality through permitting the subject who observes to be all-seeing \cite{Laval12}. However, in modern society, with the proliferation of security cameras, public concerns about privacy invasion have grown significantly. Our work provides the public a glimpse into the capability of current AI-enhanced video analysis techniques, fostering a better understanding of these technologies. 

    \item Due to the social nature, humans have a fundamental need to communicate with others. Beyond verbal languages, the choices of clothing and body postures, referred to as \emph{body languages} in this study, serve as effective means of self-expression. As illustrated in Fig. \ref{fig:style}, these non-verbal cues are particularly significant for individual seeking to establish self-esteem and personal identity. Especially for young people, how they present themselves is often a silent social signal of communicating their personality, mood and values \cite{Hethorn94}. In modern smart cities, such social signals are frequently perceived and recorded by the widespread public cameras. Therefore, it is intriguing to explore whether camera systems can act as intermediaries to convey these social signals to like-minded individuals, thereby potentially enhancing in-person relationships and prompting diverse form of self-expression.
    
    \item We also draw inspiration from a social news story in China, where a woman in Beijing occasionally found her husband on a live video broadcast at West Lake (Xi Hu) in Hangzhou. Then she instructed him to make a \emph{sign of love} towards the video camera installed lakeside, creating a romantic memory. This incident suggests that security cameras in public spaces can be repurposed as a novel communication channel to spread love and kindness. Furthermore, we speculate that preforming body languages in public spaces just as a practice of public speaking may be beneficial to supercharge players' presence and self-confidence. Thereby, it is also worthwhile to investigate the players' experiences in performing body languages in front of public cameras.

\end{itemize}
 
Owing to the above considerations, this study aims to address two questions.

\begin{itemize}
    \item[Q1:] How to gamify a public video surveillance system into a social LBG for encouraging people to engage in body language-based interactions?
    \item[Q2:] How does the social LBG impact the participants' in-person relationships in real world?
\end{itemize}

\begin{figure}[htbp]
  \centering
  \includegraphics[width=0.9\linewidth]{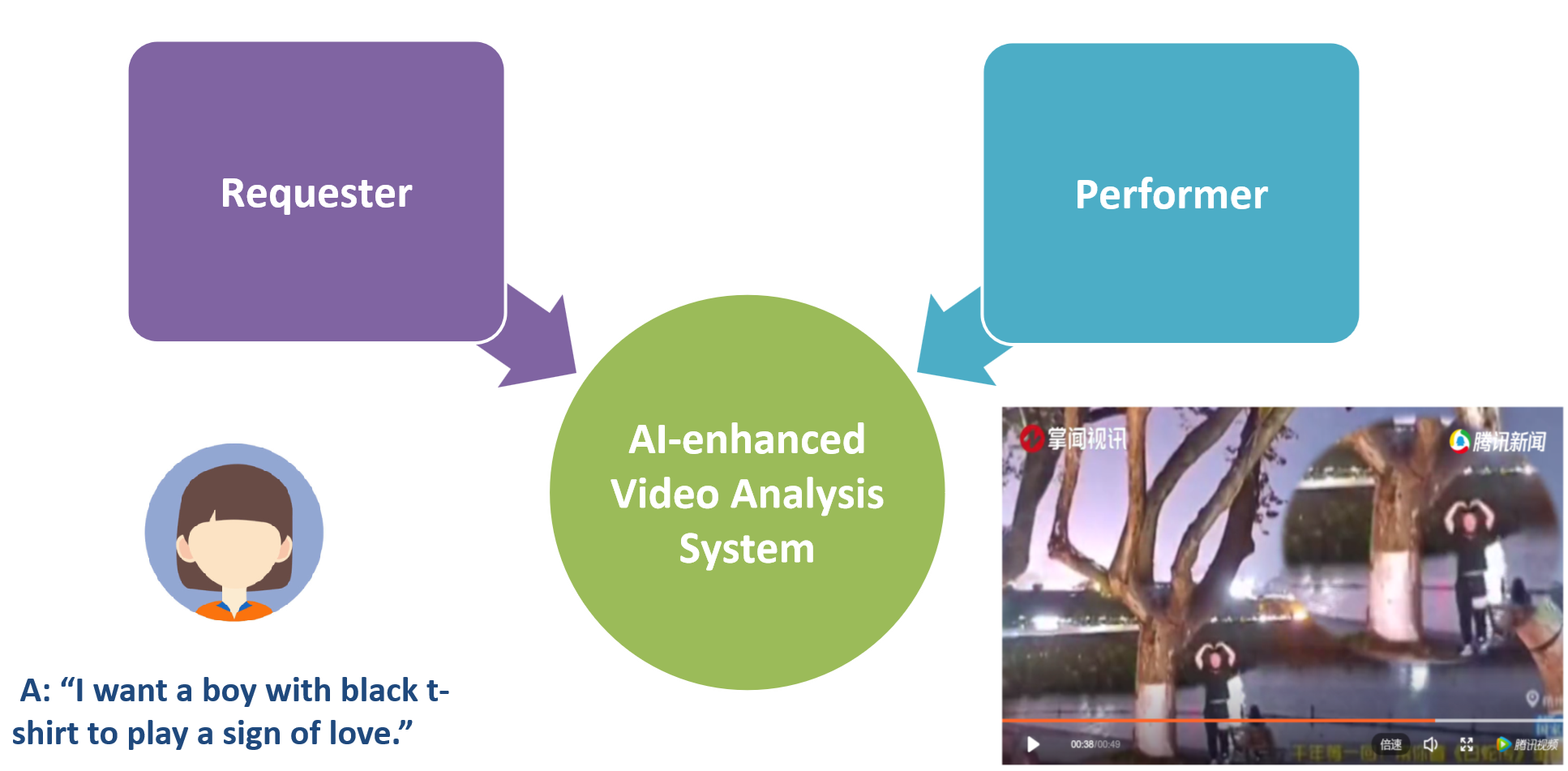}
  \caption{Basic game mechanics in LIA. The right figure is from an online news, where a woman in Beijing instructed her husband to display a \emph{sign of love} towards a camera at the lakeside of XiHu (West Lake) during an video live broadcast.}
  \label{fig:LIA}
\end{figure}
For Q1, we create a social LBG named \textbf{Love in Action} (LIA), designed to foster in-person friendships in the real world using public video cameras. As shown in Fig. \ref{fig:LIA}, players in LIA alternate between two roles: \emph{requester} and \emph{performer}. The \emph{Requester} proposes a request for a body language performance which can be customized with specific actor styles, such as "\emph{I want a boy wearing a black T-shirt to play a sign of love}". Alternatively, as a \emph{performer}, the player responds to requests by acting out the corresponding body language in front of a video camera in public spaces. To assess whether the \emph{performer}'s body language meets the \emph{requester}'s requirements, an AI-enhanced video analysis system mediates communications between players. Additionally, a WeChat mini-program \cite{Wechat} serves as the user interface, guiding players in participating in the social LBG and interacting with the AI system.

For Q2, a well-designed field study was conducted to examine the influence of LIA. Twenty-seven graduate students and staff members were recruited to experience LIA for two weeks on a technology institute's campus. Two-phase self-reported questionnaires are used to assess LIA's effect on players' in-person relationships. The results, analyzed with paired \emph{t}-test, show significantly improvements in social relationships. Furthermore, additional subjective questionnaires are performed to investigate players' attitudes towards the novel communication channel based on public video cameras. The findings reveal that players generally enjoyed the gaming experience and the new interaction style. Performing body languages in public spaces may excite players' embodied senses in in-person communications, like public speaking, potentially contributing to their social and mental well-being.

In summary, we design a social LBG, termed LIA, to encourage individuals to practice body language-based social interactions using public video cameras. A field study is performed to investigate the gaming experiences and to validate LIA's positive influence on fostering real-world social relationships.

\section{Related Work}
In this section, we review related work on LBGs designed for meaningful social interactions, recent technical advancements in intelligent video surveillance (IVS), and studies exploring the social impacts of IVS.

\subsection{Location Based Games for Meaningful Social Interactions}
LBGs are a form of pervasive augmented-reality (AR) gaming where players interact with their physical surroundings via their geographic location. The integration of global positioning system (GPS) technology in mobile devices has enabled the development of various LBGs. For example, Pokémon GO \cite{pokemon16} is one of the most well-known LBGs. Other examples include Ingress \cite{ingress}, BotFighters \cite{botfighter02}, and Geocaching \cite{geocaching05}, etc. LBGs expand traditional game boundaries by incorporating real-world landmarks, thereby surrounding players may be attracted together to compete or collaborate in social exchanges, e.g., collecting fictional creatures \cite{pokemon16} or destroying enemy's portals \cite{ingress}. Thus, LBGs provide an opportunity for people to socialize and develop real-world relationships, fostering community and friendships among players. The benefits of LBGs, e.g., Pokémon GO \cite{pokemon16}, have been recognized in enhancing users’ psychological well-being \cite{peterson21}, alleviating symptoms of depression \cite{cheng22} and promoting social interaction \cite{bhattacharya19} \cite{pokemon17}. For instance, research \cite{pokemon17} indicates that playing Pokémon GO can bring positive social experiences, where players, even strangers in surroundings, may be connected to have social conversations in the game.

Besides the aforementioned commercial LBGs, researchers have also developed various location-based serious games to prompting social relationships and mental health, such as \cite{Birn14} \cite{Xavier21} \cite{LBN22} \cite{Papangelis17}. For instance, \textit{MobileQuiz} \cite{Birn14} is designed to enhance older players' physical mobility and cognitive skills by answering a series of quiz questions while traversing different locations. \textit{CityConqueror} \cite{Papangelis17}, a multiplayer competitive LBG focused on territorial conquest based on players' locations, has an explicit ‘team-match’ setting to enhance interactions and dynamics among players within and across teams. This game has been adopted to investigate how players coordinate their movements and explore urban spaces as a group. Hirsch et al.\cite{LBN22} create a location-Based social network called \textit{Walk in My Shoes} (WIMS) encouraging users to share social content (e.g., photos) on an augmented reality (AR) map and explore new areas, thereby increasing users’ socio-spatial connection with their surroundings. 

Different from the above LBGs, the proposed LIA adopts a novel communication channel, i.e., human body language, to express love and kindness through public cameras. LIA incorporates an AI-enhanced surveillance system to evaluate the quality of body language performances. The following section discusses advanced technologies on IVS and studies on the social impacts of IVS.

\subsection{Intelligent Video Surveillance and Its Social Impacts}
Currently, video surveillance systems have been widely deployed for smart and safe cities across the world \cite{Laufs20}. To efficiently process the massive amounts of video data, significant progresses have been achieved in the filed of intelligent video surveillance. Ramachandran et al. \cite{Ramachandran12} adopt a systemic overview and comprehensive discussion on multi-camera networks, which are essential for developing \textit{situation awareness} systems. Nikoueia et al. \cite{I-ViSE20} propose an Interactive Video Surveillance as an Edge service (I-ViSE) system, comprising multiple micro-services that compute various features such as objects, colors, and behaviors, so that users are able to search for content of interest using different keywords and feature descriptions. Li et al. \cite{Li_ISEE} develop an intelligent video parsing platform based on distributed and parallel computing architectures. These platform integrates multiple visual analysis modules for efficient large-scale video parsing and person retrieval. In the HCI field, Saito et al. \cite{Saito16} build a collaborative video surveillance system which utilizes crowd-sourcing for the verification of suspicious behaviors. 

Beyond criminal control, video surveillance also serves to monitor the well-being or health care of individuals. For example, Vaziri et al. \cite{Vaziri17} develop a video game-based system (exercise gaming) named \emph{iStoppFalls} for older adults, designed to strengthen key muscles as a fall prevention intervention. In \cite{Shu21}, researchers create a multi-camera house-wide fall detection system capable of identifying various falls, including stumbling, slipping, and fainting, event from a distance. For physiological signals, Hurter and McDuff \cite{Hurter17} present \textit{Cardiolens}, a system that aids in remote heart rate monitoring. Using motion analysis and infrared imagery technologies, Li et al. \cite{Li17} present a non-contact cardiopulmonary monitoring system for detecting apnoea during sleep.

As discussed, surveillance is always 'Janus-faced', involving both control and care \cite{Lyon01}. Since the increasing rise of video surveillance post-9/11, many researchers and artists have investigated the societal impacts of surveillance from various academic and artistic perspectives \cite{Finn12}. For example, the Surveillance Camera Players (SCP) have performed a series of plays in front of surveillance cameras in New York City, which aims to criticize how the surveillance society reify a culture of public conformity and a dangerous homogeneity of behavioral display \cite{Schienke03}. Similarly, artist David Rokeby's media installation "Sorting Daemon" \cite{rokeby03} in Toronto adopt a surveillance system composed of a computer, a LCD screen and a camera, to extract image patches of pedestrian walking in a public street, then the person patches are separated into coloured regions which are sorted and reassembled into composite pictures for the display. Kihara et al. \cite{kihara19} design a critical pervasive game based on surveillance cameras with an escape room theme, using surveillance cameras to raise public awareness about the socio-technical implications of AI-enhanced urban surveillance. In contrast, Albrechtslund and Dubbeld \cite{Albrechtslund05} propose an positive view of surveillance, suggesting it can be protective as well as playful and enjoyable. The authors discuss surveillance games, such as “Can You See Me Now?” \cite{Anastasi02} and “Monopoly Live” \cite{Ogles05} which are early LBGs played outdoors using GPS on mobile devices. Taylor \cite{Taylor15} also explores the concept of “play to the camera", examineing how cameras transform ethnographic fieldwork and e-sports performances. Compared to previous surveillance games, in this paper, we explore the use of public surveillance cameras for players to engage in body language-based social interactions, offering a unique perspective on surveillance and social interactions.

\section{System Design}
This section introduces the essential elements in designing LIA, including the game mechanics, the AI-enhanced video analysis system that mediates the communications among players, and the user interface.

\subsection{Game Mechanics}
\begin{figure}[h]
  \centering
  \includegraphics[width=1\linewidth]{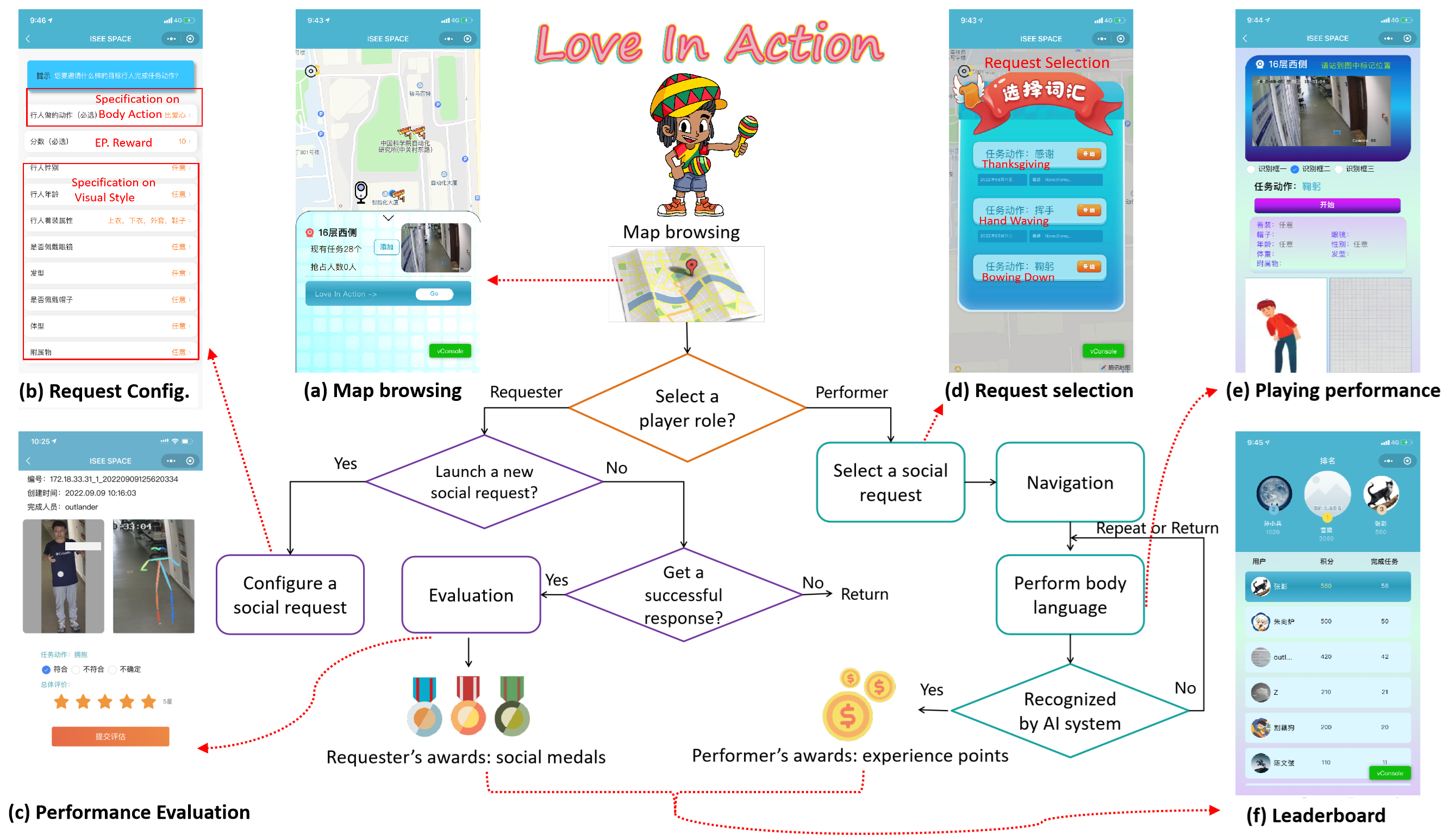}
  \caption{The flowchart of LIA gaming process.}
  \label{fig:flowchart}
\end{figure}

The flowchart of the LIA game mechanic is shown in Fig. \ref{fig:flowchart}. Like other mobile LBGs, e.g., Ingress and Pokemon GO, LIA offers an Augmented Reality (AR) map for users to explore nearby areas. The AR map visually depicts the locations of surrounding public video cameras and displays information about social requests associated with each camera. Player can use this information to select their role, either \textit{requester} or \textit{performer}.

\begin{itemize}
    \item A \emph{requester} can initiate a new social request, offering certain rewards that may require players to spend some \emph{experience points} (EPs) in the game. The request must be customized with specific actor styles, including the visual style (attributes) of the desired performer, and the action the performer should execute. Once a \emph{performer} successfully responds to the social request, the \emph{requester} is prompted to review and evaluate the body language performance. Based on this evaluation, a \emph{social medal} will be granted as the reward for the \emph{requester}.
    
    \item A \emph{performer} should first review the existing social requests in terms of their configurations (desired visual style and action). They then select a suitable request to fulfill and perform the specified body language. Guided by mobile GPS navigation, the \emph{performer} goes to the target location and acts out the specified body language in front of a camera. If the AI evaluator determines that the performance is qualified to fit the request's configuration, the \emph{performer} will be rewarded with the EPs promised by the \emph{requester}. Then, the player can use these EPs to initiate new requests, thereby incentivizing continued participating.
\end{itemize}

Since WeChat, akin to WhatsApp, is the most popular social media platform in China, we have developed the user interfaces of the LIA as a WeChat mini-program \cite{Wechat}. The mini-program is a lightweight application within the WeChat ecosystem, leveraging its fast loading speeds and comprehensive social features, such as Friendship links and Moments etc. Six main interface pages are presented in Fig. \ref{fig:flowchart}, corresponding to different scenarios in the game. The page (a), presents the AR map interface, enabling users to browse and navigate to specific camera view. The page (b) allows the \textit{requester} to initiate a request by specifying certain actions and attributes as configurations, with options including five body actions (\textit{love sign, bowing down, thanksgiving, hand waving and hugging}) and various attribute categories (e.g,, \textit{gender, clothing types}). The page (c) is for user evaluation of received performance, where the \textit{requester} evaluates the visual style and body action with an overall score. The page (d) is for the \textit{performer} to select a social request. The page (e) displays the real-time detection result in the camera when a \textit{performer} is executing a body language performance. The final page (f) is a leaderboard ranking players by the number of social medals they have obtained and the EPs.

In summary, the reward system in LIA includes two types of rewards: \emph{experience points} (EPs) and \emph{social medals}. EPs are earned by \emph{performers} for fulfilling \emph{requesters}’ social requests, which should be paid by \emph{requesters} to issue their own requests. Meanwhile, when a \emph{requester}'s request is successfully completed, they will be awarded a \emph{social medal}. The more requests a player receives that are fulfilled by others, the more \emph{social medals} they accumulate. In the game's leaderboard design, \emph{social medals} are prioritized over EPs in player rankings. This setup is expected to encourage players to spend their EPs to elicit responses to their requests, thereby sustaining the economic system within LIA.

\subsection{AI-enhanced Video Analysis System}
In this study, an AI-enhanced video analysis system is built to mediate communication among players by assessing whether the performer's visual style and body actions satisfy the requester's requirements. The system comprises four video analysis modules: \textit{pedestrian detection $\&$ tracking}, \textit{human attribute recognition}, \textit{pose estimation} and \textit{skeleton-based action recognition}, which are integrated into a scalable and composable video parsing platform \cite{Li_ISEE}.

\begin{figure}[htb]
  \centering
 \includegraphics[width=\linewidth]{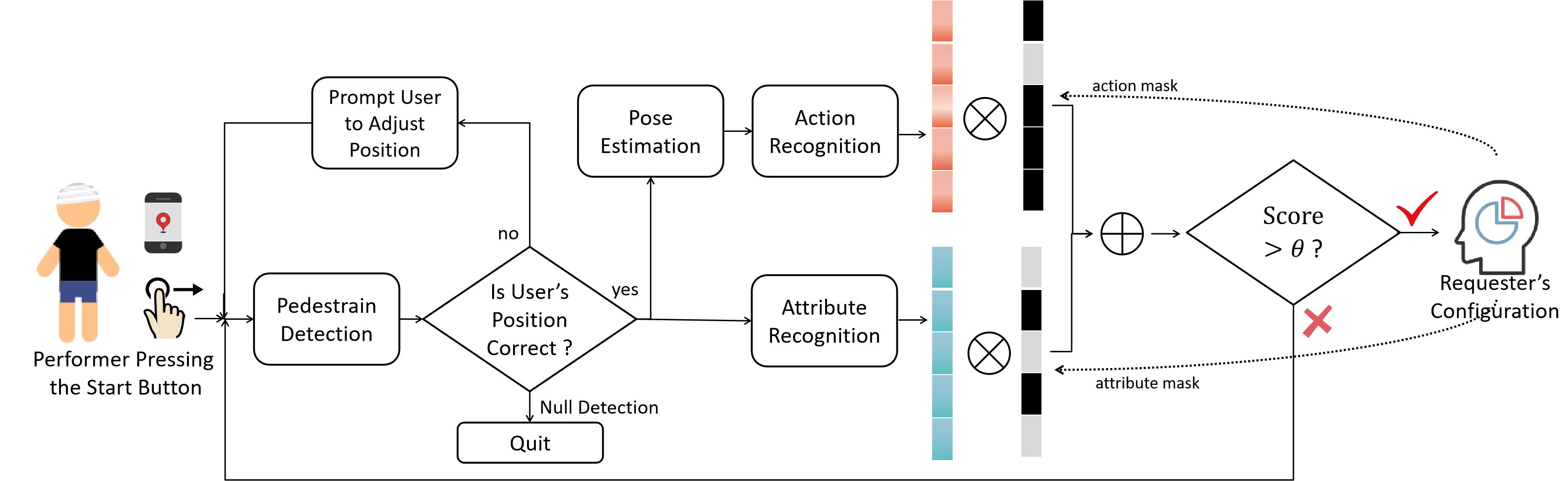}
  \caption{The pipeline of performance quality evaluation, where $\bigotimes$ denotes the element-wise product between two vectors and $\bigoplus$ denotes the weighted average calculator.}
  \label{fig:evaluation}
\end{figure}

As presented in Fig. \ref{fig:evaluation}, upon arriving at the camera scene in response to a social request, the performer must first press a start button via the WeChat Mini-program to signal their readiness to perform. This action prompts the AI system to verify the performer’s position using mobile GPS and pedestrian detection, where the YOLOv5 model \cite{yolov5} pre-trained on the COCO dataset serves as the pedestrian detector. Subsequently, the performer’s image undergoes attribute recognition using the DeepMar model \cite{li2015deepmar} pre-trained on the RAP dataset \cite{2019rap}. For action recognition, the pre-trained AlphaPose \cite{fang2017rmpe} and VideoPose \cite{videopose2019} models estimate the 3D joints of the detected person in each image frame, which are then input into the EfficientGCN model \cite{EfficientGCN} to identify the action.

The AI system evaluates a performer's response to a social request through comparsing their body language performance against the request. As shown in Fig. \ref{fig:evaluation}, the action and attribute recognition results (indicated by the red and blue vectors) are combined with the action and attribute masks through an element-wise product to yield the final recognition results associated with the social request. Then, a score-level fusion process calculates a matching score. 

Specially, let $P^m\in R^{K}$ represent the predicted probabilities for $K$ action categories and the probabilities $P^a\in R^{L}$ for $L$ attribute categories. For a given social request configured as $<k, S>$, where $k$ is the index of the required action and $S\subset\{1,2,...L\}$ is the set of relevant attribute indices, the matching score is calculated as follows.

\begin{equation}
    score = \alpha\sum_{i=1}^{K}\textbf{1}(i=k)\log p_i^m +\frac{(1-\alpha)}{|S|}\sum_{j=1}^{L}\textbf{1}(j\in S)\log p_j^a,
    \label{Eq:score}
\end{equation}
where the function $\textbf{1}(.)$ selects the action and attribute recognition results corresponding to the configuration $<k, S>$. The weight $\alpha$ represents user preference for body action or visual style. In this work, we set $\alpha=0.7$ in all experiments. Noted that attribute recognition is typically formulated as a multi-label classification problem with multiple binary attribute categories, allowing for the selection of multiple pedestrian attributes to define the social request. 

Finally, the matching score is compared against a pre-defined threshold $\theta$ to determine whether the performance satisfies the social request.

\section{Field Study}
\subsection{Game Environment}
\begin{figure}[htb]
  \centering
 \includegraphics[width=0.9\linewidth]{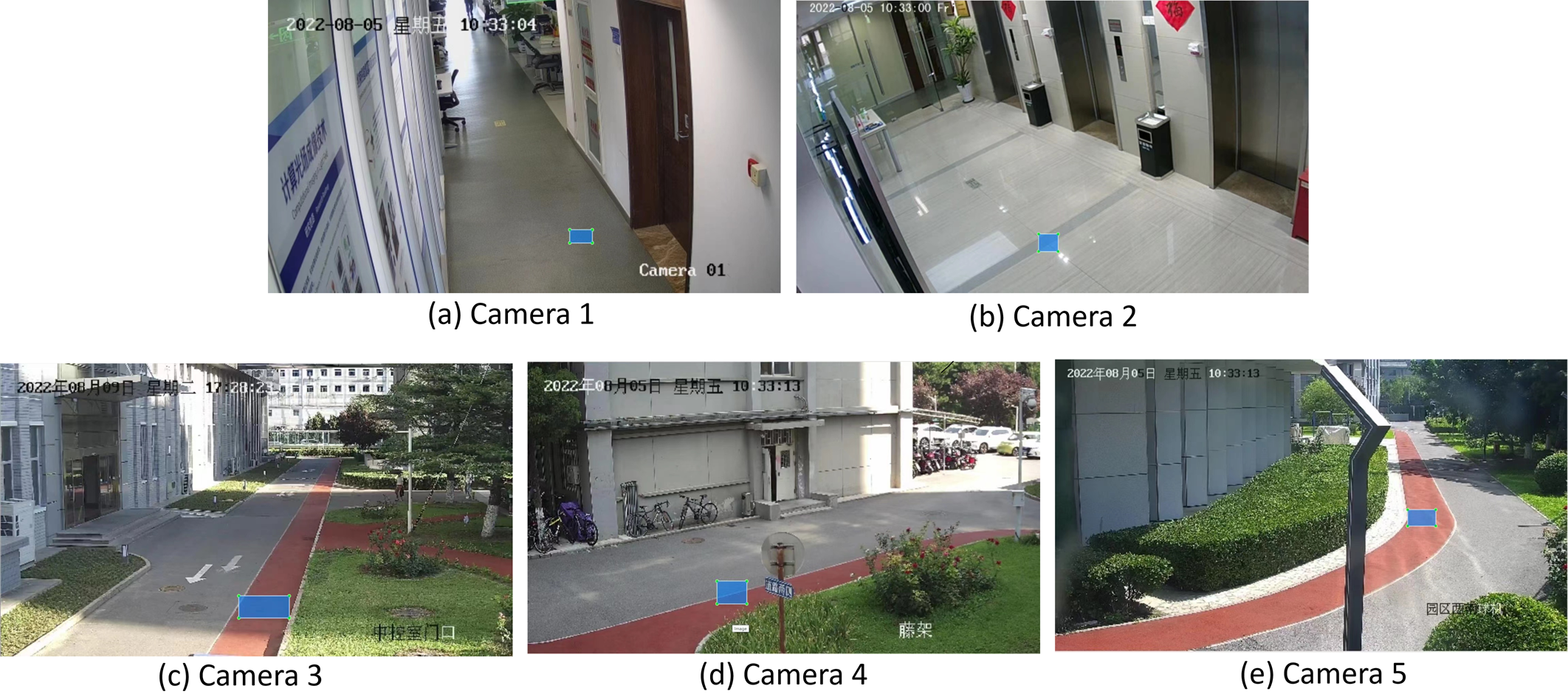}
  \caption{Illustrations of the five camera scenes selected for experiencing the LIA.}
  \label{fig:cameras}
\end{figure}

Due to the sensitivity of the use of public camera data, our filed study is conducted within a limited and controlled game environment. As shown in Fig. \ref{fig:cameras}, the field study utilizes five video cameras from an institute campus, with three installed in outdoor scenes and two in indoor areas - a corridor and an elevator lobby. The blue areas in the scenes indicate the pedestrian detection zones, used to identify the \textit{performers}.

It should be noted that the access to these public cameras is supported by a National Key Research and Development Program of China, with the objective of developing advanced AI techniques for situation sensing in large-scale scenes. All video/image data is meticulously preserved and only used within the limited duration of the field study.

\subsection{Participants}
To assess the LIA's impact on interpersonal relationships, we recruit 27 participants from a middle-size WeChat group with total 181 members, all affiliate with a research group at the institute. Demographic data, including gender, age, job position, and WeChat ID, are collected prior to the game. The participant pool comprises 9 females and 18 males with an average age of 28.6 years (SD = 5.2). In terms of job positions, there are 9 graduate students, 14 technicians and 4 research staff.

As all participants belong to the same research group, there have been pre-existing social ties among them. Before the game, we ask about the number of their WeChat friends participating in the LIA and the nature of these relationships (close/non-close). In summary, Of the 25 respondents, 62.8$\%$ of pairwise relationships (424 out of 675) are pre-existing WeChat friendships. However, the majority of these friendships (71.7$\%$, 304 out of 424) are considered non-close, indicating a lack of socio-emotional communication in their daily interactions. All participants provide consent for data collection and processing in accordance with the Personal Information Security Specification (PISS) in China \cite{PISS}.

\subsection{Procedures}
To examine the influence of the LIA game on the social relationships among participants, a comparative study is conducted. The study includes two phases. For the first phase, participants are required to complete a \textit{Social Relationships Index} (SRI) questionnaire \cite{sri09} prior to gameplay. This questionaire aims to assess their perceived friendships with six WeChat friendship contacts, self-selected from all participants. For each friend, participants respond to the same six questions using a 6-point scale (ranging from 1 = not at all to 6 = extremely), rating their subjective impressions of the friendships. Furthermore, participants are asked to categorize each of the six friendships as either 'close' or 'non-close'. The details of the SRI can be found in \cite{sri09}.

After introducing the LIA game's context and rules, all participants are allocated 100 EPs to publish their initial social requests, making the commencement of the game. Players can play the game with the WeChat Mini-program, conveniently fitting gameplay into their daily work or study routines. The complete gameplay records for each participants are preserved for subsequent analysis. Approximately two weeks later, in the second phase, participants are again asked to complete the SRI questionnaire for the same six friends. 

Moreover, a brief review including three types of subjective questions is administered to gauge participants’ special experiences when playing the LIA. The response options for these subjective questions are scaled from 1 (not at all) and 5 (extremely).
\begin{itemize}
\item[(1)] \textbf{Performer's Embodied Sense.} Participants are required to assess whether they experience an embodied sense similar to public speaking or direct face-to-face communication when performing body language in public spaces.
\item[(2)] \textbf{Requester's Social Satisfaction.} Participants are queried about the level of satisfaction and happiness they derived from the LIA game, particularly when reviewing the body languages exhibited by other participants.
\item[(3)] \textbf{Users' Attitudes to AI-mediated Communication.} Participants are also asked to rate the effectiveness of the AI system during their experiencing with the LIA.
\end{itemize}

Finally, descriptive statistical analysis is conducted utilizing personal gameplay records, SRI questionnaire answers and subjective review results.

\section{Results}
In this section, we report the statistical findings from the field study, and qualitative analysis derived from the questionnaires and brief reviews.

\subsection{Overall Statistics}
Over approximate two weeks, 25 active participants (those who perform at least one body language) complete a total 411 body language performances, with 76.9$\%$ passing the AI system's quality evaluation. Additionally, 11 users obtain more than 10 social medals. Following the final review and evaluation by the \textit{requesters}, 268 social medals are granted to 23 players, with the top player receiveing 31 social medals, indicating that the player has 31 social requests completed by others.

\begin{figure}[htb]
  \centering
 \includegraphics[width=\linewidth]{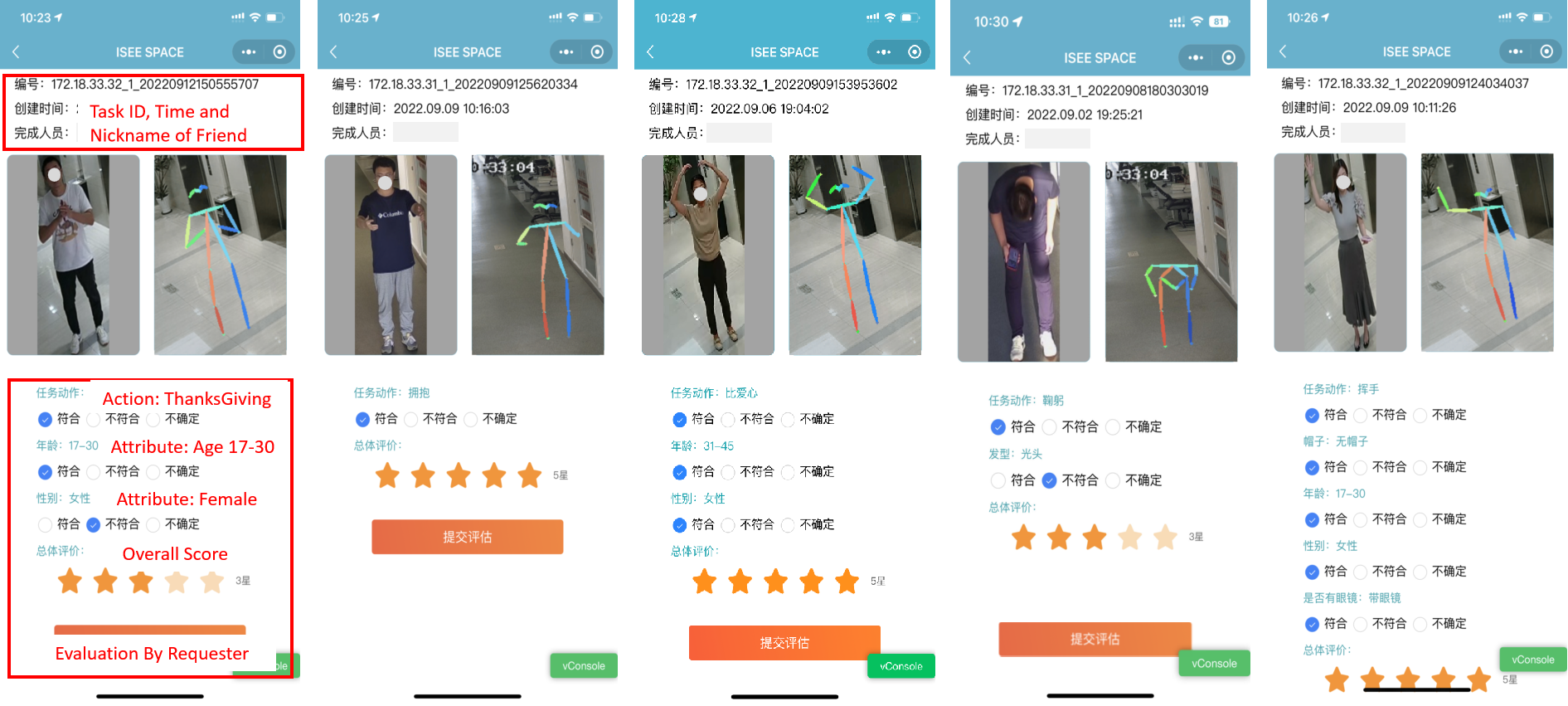}
  \caption{Examples of user's evaluations on body languages, where the five kindness actions are \textit{thanksgiving}, \textit{hugging}, \textit{sign of love}, \textit{bowing down} and \textit{hand waving}}
  \label{fig:instances}
\end{figure}

At the final review and evaluation phase, users who publish social requests are required to provide an overall score (on a scale of 1 to 5) for the received performance and confirm whether the visual style (attributes) and body actions match their social requests. Fig. \ref{fig:instances} shows five examples of such evaluation process for five action categories. Across all user evaluations, the overall score statistics reveal a mean value of 4.19 and a standard deviation (SD) of 1.13. Additionally, 65.9$\%$ of attributes are confirmed to align with the visual style requirements, and 90.3$\%$ body actions are consistent with the expectations of \textit{requester}. The high precision (90.3$\%$) in action recognition demonstrates the effectiveness of the AI-based performance evaluation. Meanwhile, there remains considerable room for improving the precision of visual style (attribute) recognition.

\subsection{Impacts on Social Relationships}
Based on the pre- and post-game SRI questionnaires \cite{sri09}, we firstly calculate the friendship positivity as the average value of the first three questions, which relate to the perceived helpfulness of friends when needing advice, understanding, or a favor in the SRI. Then, we adopt the mean and SD values of the total 150 friendships (6 friendships per active player) to measure the divergence between the pre- and post-game distributions of friendship positivity. Furthermore, according to the division of close/non-close friendships (58/92) in the pre-game questionnaire, we also investigate the influences on these two types of friendships, respectively.

\begin{figure}[thb]
  \centering
 \includegraphics[width=0.7\linewidth]{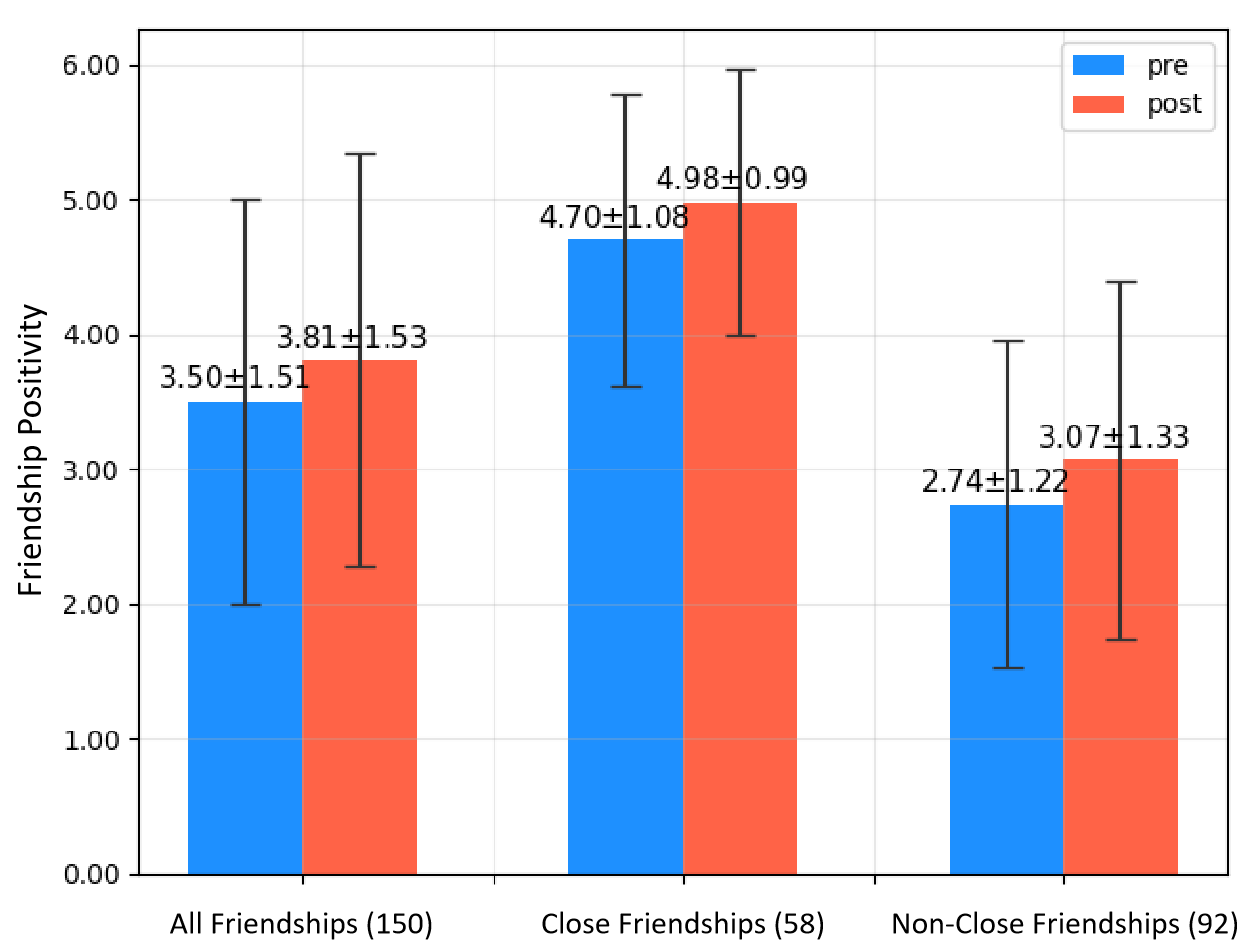}
  \caption{Influences of the LIA on social friendships. The values in the parenthesis, i.e., 150, 58 and 92, denote the numbers of friendships in different groups.}
  \label{fig:f}
\end{figure}

Fig. \ref{fig:f} shows the comparative results of friendship positivity for the entire group, as well as for the two types of friendships. The results indicate a significant increase in positivity across all three groups post-LIA. Notably, the improvement in non-close friendships is larger than that of close friendships. The observation might be attributed to the lower baseline score of non-close friendships, making them more susceptible to enhancement through kind actions by others. Furthermore, we conduct a paired \emph{t}-test to measure the statistical significance of the improvements in social friendships. 

\begin{table}[htb]
  \centering
  \caption{Results of paired \textit{t}-test measuring the significance of the improvements on social friendships.}
  \label{tab:t_test}
  \begin{tabular}{cccc}
  \toprule
     & All Friendships & Close Friendships & Non-Close Friendships\\
  \midrule
    \textit{t}-values & 6.68 & 4.17 & 5.23 \\
    \textit{df} & 149 & 57 & 91 \\
    $t_0$ (0.001) & 3.15 & 3.24 & 3.18\\
  \bottomrule
\end{tabular}
\end{table}

Table \ref{tab:t_test} displays the \textit{t}-values for the three groups of friendships. It is evident that all three \textit{t}-values exceed the corresponding \textit{t0} with a one-tailed \textit{p-value} of 0.001 at varying degrees of freedom (\textit{df}). Consequently, the improvements in all types of friendships are statistically significant at the 0.001 level, as indicated by the t-test. Our findings demonstrate the positive impact of the LIA on social cohesion within local communities, aligning with previous studies that explore the fostering of meaningful social interactions through well-designed LBGs \cite{Xavier21} \cite{Xavier20} and Pokémon Go \cite{bhattacharya19}. However, unlike prior studies which aims to engage players together for face-to-face social interactions in physical space, the LIA employs an AI-mediated communication system to facilitate asynchronous interactions between friends. Furthermore, the LIA utilizes a novel body language-based communication channel to convey kindness through video cameras in public spaces. The next section will discuss these special user experiences within the LIA.

\subsection{User Experiences}
Overall, the participants feel excited when performing body languages in public spaces and enjoy the asynchronous social interactions with others in the LIA. As shown in Section 4.4, three subjective questions on \textit{performer's embodied sense}, \textit{requester's social satisfaction} and \textit{user's attitudes to AI-mediated communication}, are asked for rating their experiences with a 1-5 scale.

\begin{table}[htb]
  \centering
  \caption{Results of user experiences on the LIA, where each question is asked to rate a 1-5 scale.}
  \label{tab:user_exp}
  \begin{tabular}{cccc}
    \toprule
     & Embodied Sense & Social Satisfaction & Attitudes to AI\\
    \midrule
    Mean & 3.44 & 4.16 & 3.76 \\
    SD & 0.98 & 0.78 & 0.99 \\
  \bottomrule
\end{tabular}
\end{table}

As depicted in Table \ref{tab:user_exp}, the statistics (Mean and SD) across 25 active participants reveal that the majority expressed satisfaction with their feelings when reviewing others' body language based on the requester's specifications. As for the performer's embodied sense (M=3.44, SD=0.98), 14 of 25 participants gave the rates larger than 3, which indicates that players experience a senses of embodied performance during the LIA, akin to practicing public speaking in front of a crowd of persons. Regarding the attributes towards AI, users had an overall positive view of AI-mediated communication (M=3.76, SD=0.99), indicating that they feel the AI-based evaluation satisfying for their experiences in LIA.

\section{Discussions}
\subsection{Impacts on Social Relationships}
As shown in Fig. \ref{fig:f} and Table \ref{tab:user_exp}, the overall post-game friendship positivity rose from 3.50 to 3.81, and the players' social satisfaction is also confirmed (4.16 out of 5). To comprehend the positive impact, we request feedback from some players regarding the underlying reasons. A representative feedback is as follows.   

"\emph{I'm a bit introverted when it comes to socializing with people. When I interact with unfamiliar colleagues, I often feel embarrassed. The game provides an interesting and indirect way to express kindness to them. I also enjoy my colleagues' body language performances. Moreover, I've noticed that some individuals also share my favorite interests, as evidenced by their clothing styles.}"

As presented in the above feedback, the positive impacts on players can be understood from two main aspects. Firstly, player's social requests are fulfilled when others perform body language for them, leading to an improvement in their social medal ranking on the leaderboard. Secondly, as requesters review the received performances, they also have the opportunity to closely observe the performer's clothing and apparel, thereby gaining insights into their personalities and values from these subtle social cues. Given that LIA players may interact in their daily work, this information serves as additional conversation topics, so as to foster better relationships among them.

We also find an interesting exception in the post-game assessment of friendship positivity. A player initially evaluates a non-close friendships with a positivity score of 2 before playing LIA, which drop to 1 after the game. To understand this phenomenon, we ask the player for the reason. She explains that the decline is due to the friend's failure to adhere to the specified visual style in her social request.

"\emph{Actually, I ask a female friend to perform a body language for me. However, the male friend does not check the condition carefully, and the AI evaluator incorrectly accepts the performance. Therefore, I lower the positivity score for this friend after the game, as he seems unreliable in fulfilling my requests.}" 

\textbf{The explanation emphasizes the critical role of AI-enhanced video analysis system in ensuring the efficacy of LIA in cultivating friendships.} In our short-term gaming experiences, the majority of participants comply with the requests to perform body language. However, as the extension of the game duration and an increase in the number of participants, a number of falsely accepted low-quality performance may damage trust and friendships between players. Thereby, it is essential to fully utilize the player data generated during the LIA to refine AI models continually.

\subsection{Embodied Experiences of Body Languages}
LIA requires participants to perform body languages in public spaces, facing a video camera. Though this may initially seem awkward due to the unusual physical movements that could surprise those people in the vicinity, it also has positive aspects. Performing body language in public can excite players, generating a stronger senses of embodiment which is often lacking in many online social media platforms. As indicated in Table \ref{tab:user_exp}, the embodied sense score of 3.44 out of 5 is moderately high, indicating a significant level of user engagement. One player described the experience with performing body language as follows.

"\emph{When learning the game, I find it somewhat awkward to make body postures in front of the security camera in the yard. However, in practice, it appears that most individuals pay little attention to my actions. Some may be curious about the specific physical movements I am making, but they don't ask me, perhaps because the performance is very brief. Nevertheless, I still experience some anxiety when performing the sign of love, a gesture rarely observed in public, unlike the more common gesture of hand-waving.}"

As described, players' concerns about embarrassment from public performance primarily exist in their imaginations, as all signs of kindness in LIA can be performed quickly, taking about 30 seconds. During LIA gaming experiences, all participants are able to complete the performance tasks smoothly, without interference from others.

\subsection{Attitudes to AI-enhanced Video Surveillance}
As shown in the third column of Table \ref{tab:user_exp}, players' attitudes toward the AI evaluator are overall satisfactory, with a score of 3.76 out 5. However, some players have provided feedback expressing confusion about how the AI system classifies their body language. Especially when their performance is classified with errors and inconsistently. For example, given a social request published at two cameras, the same performance of one player is accepted by \emph{Camera A}, while denied by \emph{Camera B}. They are confused by the classifier's inconsistent responses under different cameras. Therefore, it is essential to solve the alignment issue between AI systems and human users for future human-AI interactions, such as by developing the explainable AI (XAI) models instead of current black-box AI systems.

Nevertheless, when playing the LIA game, participants become much more aware of the prevalence of security cameras in their daily lives than they were before. Some players also express strong concerns about privacy and safety issues related to the uses of video cameras and the LIA mini-program. In response to these concerns, we explain the three design principles of LIA to all players. 

\begin{itemize}
\item[(1)] Firstly, LIA is designed to strengthen existing WeChat friendships, rather than initiate new relationships. Players can only interact with and respond to social requests from their WeChat friends. This ensures that the dissemination of their activity pictures during the LIA gaming session is managed.
\item[(2)] Regarding safety, players might try to deliberately deceive the AI evaluator by displaying inappropriate or indecent postures in response to social requests. However, the public nature of body language performance in public areas, along with the video clips recorded by the surveillance system, will effectively deter such attempts. 
\item[(3)] Previous studies have raised various criticism and questions regarding the widespread use of surveillance technologies \cite{Schienke03} \cite{kihara19}. However, surveillance technology is inherently two-faced, embodying both “care-and-control” aspects \cite{Lyon01}. In LIA, we aim to seriously consider the positive aspects of AI-enhanced surveillance systems to investigate its potential in enhancing users' well-beings, which offers a more balanced perspective on surveillance technologies.
\end{itemize}

\subsection{Limitations}
From the above analyses and discussions, there are still some limitations in the preliminary work. Firstly, the scale of participants and the duration of the field study are limited (27 players and 2 weeks), which prevent us from drawing more comprehensive and convincing conclusions. Secondly, the capabilities of AI models within LIA are still limited, characterized by imperfect recognition of human attributes and an inability to provide explanations, which may hinder the development of closer social relationships among users and the establishment of trust in AI models. Finally, the social requests in the current LIA are limited to a pre-defined vocabulary of human attributes and body postures, which restricts players from expressing individual preferences for more diverse body languages, such as the exampled styles in Fig. \ref{fig:style}. 

To address the first issue, considering privacy and safety, perhaps we could deploy a long-term user study in a closed and controlled environment, such as a public recreation area, positioning it as a leisure game for office workers. Regarding the last two issues, recent progress in large vision-language foundation models present promising solutions for more accurate body language recognition and more natural human-AI interactions.

\section{Conclusion}
In this work, we design and create a social game, namely LIA, which engage people in practicing body language in public spaces to fulfill others' social requests through video cameras. We built an AI-enhanced video analysis system that integrates multiple visual analysis modules to support body language-based communications among users. A two-week field study is conducted to investigate the influence of LIA on fostering in-person friendships in the real world. The comparative study results demonstrate improvements of social relationships in the LIA gaming experiences among participants.

\section{Acknowledgement}
This work is supported in part by the National Key Research and Development Program of China under Grant 2016YFB1001005, in part by the National Natural Science Foundation of China (62373355, 62306311). Zhang Zhang would also like to thank Jingyao Zhang, Jingxuan Zhang and Xiangyi Wang, for their helpful discussions and kind support in the design of the LIA game mechanism.

%
%
%
\bibliographystyle{splncs04}
\bibliography{mybib}

\end{document}